\documentclass[notitlepage, aps, prb, 10pt, superscriptaddress, twocolumn]{revtex4-1}
\pdfoutput=1

\usepackage{mathptmx}
\usepackage{amsmath}
\usepackage{bm}
\usepackage{graphicx}

\usepackage{amsfonts}
\usepackage{amssymb}
\usepackage{amsthm}
\usepackage{mathtools} 
\usepackage{acronym}
\usepackage[normalem]{ulem}
\usepackage{siunitx}
\usepackage{color}
\usepackage{todo}
\usepackage{xfrac}
\usepackage{tabularx}

\newcommand{\figwidth}{3.375in}

\newcommand{\new}[1]{\textcolor{black}{#1}}

\begin{document}

\title{Role of longitudinal fluctuations in L1$_0$ FePt}

\author{Matthew O. A. Ellis}
\affiliation{School of Physics and CRANN, Trinity College Dublin, Dublin 2, Ireland}

\author{Mario Galante}
\affiliation{School of Physics and CRANN, Trinity College Dublin, Dublin 2, Ireland}

\author{Stefano Sanvito}
\affiliation{School of Physics and CRANN, Trinity College Dublin, Dublin 2, Ireland}

\begin{abstract}

L$1_0$ FePt is a technologically important material for a range of novel data storage applications.
In the ordered FePt structure the normally non-magnetic Pt ion acquires a magnetic moment, which
depends on the local field originating from the neighboring Fe atoms. In this work a model of FePt is
constructed, where the induced Pt moment is simulated by using combined longitudinal and rotational
spin dynamics. The model is parameterized to include a linear variation of the moment with the exchange
field, so that at the Pt site the magnetic moment depends on the Fe ordering. The Curie temperature of FePt
is calculated and agrees well with similar models that incorporate the Pt dynamics through an effective Fe-only
Hamiltonian. By computing the dynamic correlation function the anisotropy field and the Gilbert damping are
extracted over a range of temperatures. The anisotropy \new{exhibits a power-law dependence with temperature
with exponent $n\approx2.1$. This agrees well with what observed experimentally and it is obtained without including 
a two-ion anisotropy term as in other approaches.} Our work shows that incorporating longitudinal fluctuations into 
spin dynamics calculations is crucial for understanding the properties of materials with induced moments. 

\end{abstract}


\maketitle

\section{Introduction}

The increasing demand for high density data storage has driven the adoption of novel storage technologies.
Heat assisted magnetic recording (HAMR) is one such technology. HAMR aims to overcome the super-paramagnetic
limit in hard disk drives media with ultra-small grain structure by using highly anisotropic magnetic materials. The
particular phase at the forefront of HAMR is L$1_0$-ordered FePt, which exhibits an anisotropy large enough to
stabilize data storage on grains only a few nanometers wide.\cite{Weller2000} Crucial to HAMR is the temperature
dependence of the magnetic anisotropy, since writing data on such materials is possible only at high temperatures,
where the anisotropy is reduced. Measurements showed that in L$1_0$ FePt the first-order anisotropy has an unusual
temperature dependence of $K(T) \propto M(T)^n$ where $n=2.1$ as opposed to $n=3$ predicted for typical uniaxial
anisotropy.\cite{Okamoto2002,Zener1954,Callen1966} Such anomalous dependence can be explained with a
two-lattice model, in which both the magnetic moment and the anisotropy are carried by two different sub-lattices.
\cite{Skomski2003}

FePt in the L$1_0$ structure forms alternating planes of Fe and Pt ions along the $c$-axis, and the large magnetic
anisotropy energy arises due to the strong spin-orbit coupling of the Pt atoms and the $d$-orbital
hybridization\cite{Okamoto2002,Daalderop1991}. As shown by previous calculations alloying induces a magnetic
moment on the normally non-magnetic Pt. In addition, the size of this moment was seen to vary linearly with the
collinearity of the neighboring Fe moments such that in a ferromagnetic configuration Pt is locally magnetic,
while in an anti-ferromagnetic one it is diamagnetic.\cite{Mryasov2005} This complexity is problematic for simulations
using the Heisenberg model, as this assumes the moments are unit vectors constant in magnitude. In order to circumvent
this issue Mryasov \emph{et al.}\cite{Mryasov2005} defined an effective Heisenberg energy, where the relation
between the Pt moment and the local field due to the Fe atoms is used to reconstruct a Hamiltonian containing only the Fe
degrees of freedom. This model has been used extensively to simulate properties of L1$_0$ FePt systems with great success\cite{Hinzke2008,Kazantseva2008PRB,Ellis2015_FePt,Ellis2016}. Within this framework the temperature dependence
of the anisotropy arises from the combination of the first-order anisotropy (giving a $n=3$ exponent) and an effective two-ion
anisotropy ($n=2$). However, Mryasov's model has the limitation that it does not directly simulate the Pt moments. Thus in
non-equilibrium situations, such as those produced by the alternating spin-transfer torques observed in FePt tunnel
junctions\cite{Galante2019} or excitations by a laser\cite{Yamamoto2018a}, the full details of the dynamics cannot be modeled.

Here we construct an alternative model of induced Pt moments in FePt, \new{incorporating longitudinal spin fluctuations into a generalized spin dynamics scheme}.
In this, the atomic magnetic moments are not considered to have a constant length but rather change dynamically.
Building upon the work of Ma \emph{et al.}\cite{Ma2012b} the Heisenberg Hamiltonian is extended to include a Landau-like
longitudinal energy term, which for Pt is set so that the Pt moment depends on the local order of the Fe atoms. Thus our Hamiltonian
is effectively a two-spin model with \new{additional} longitudinal fluctuation. \new{We find that this model can correctly reproduce the 
$n=2.1$ exponent observed for the temperature dependence of the anisotropy, without the need to introduce explicitly any mediated 
two-ion anisotropy term.} The rest of the paper is arranged as follows. Firstly the methodology
describing the longitudinal model of FePt is detailed including a description of Mryasov's model. Then, we present our results
on the temperature dependence of various magnetic properties. These include the magneto-crystalline anisotropy calculated
from a ferromagnetic-resonance-type experiment. Finally we present our conclusions.

\section{Methodology}

The pioneering work of Landau and Lifshitz\cite{Landau1935}, and later Gilbert\cite{Gilbert2004}, presented an equation of
motion for a magnetic moment, which has been the corner-stone for the numerical modeling of magnetic materials for many
years\cite{Ellis2015_LL,Evans2014}. The Landau-Lifshitz-Gilbert (LLG) equation, as it is commonly referred, describes the
transverse rotational motion of the magnetization. The dynamics is driven by precessional and damping terms so that the
longitudinal length of the moment is conserved. The LLG equation in terms of atomic spin moments, $\mathbf{S}_i$, takes the form

\begin{equation}
  \frac{\partial \mathbf{S}_i}{\partial t} = -\gamma \mathbf{S}_i \times \mathbf{H}_i + \lambda  \mathbf{S}_i \times \frac{\partial \mathbf{S}_i}{\partial t}, \label{eq:LLG}
\end{equation}

where $\gamma = \SI{1.76e11}{s^{-1}T^{-1}}$ is the gyromagnetic ratio and $\lambda$ is the atomistic damping parameter,
which is the limit of the Gilbert damping at zero temperature. If $\mu_i$ is the equilibrium magnetic moment of each atom
taken as the normalization constant, the spin vector entering the dynamics will be defined as $\mathbf{S}_i = \mathbf{m}_i/ \mu_i$. 
%
%
In Eq.~(\ref{eq:LLG}) $\mathbf{H}_i = (1/\mu_i) \partial \mathcal{H} / \partial \mathbf{S}_i + \boldsymbol{\xi}_i$ is the effective
magnetic field acting on the $i$-th spin, which comprises a term arising from the spin Hamiltonian, $\mathcal{H}$, and a fluctuating
thermal noise, $\boldsymbol{\xi}$. Conventionally the extended Heisenberg Hamiltonian is used, which reads
\begin{equation}
  \mathcal{H} = - \sum_{i, j \neq i} J_{ij} \mathbf{S}_i \cdot \mathbf{S}_j - \sum_i k_i (\mathbf{S}_i \cdot \mathbf{\hat{e}}_i )^2 - \sum_i \mu_i \mathbf{S}_i \cdot \mathbf{H}_\text{app}, \label{eq:HHam}
\end{equation}
where $J_{ij}$ is the exchange coupling between the spins $i$ and $j$, $k_i$ is the onsite uniaxial anisotropy along the
axis $\mathbf{\hat{e}}_i$ and $\mathbf{H}_\text{app}$ is the external applied field. 
\new{In general, the uniaxial anisotropy constant can contain various contributions, but in many cases it is the magneto-crystalline 
anisotropy (MCA), arising from the quantum mechanical spin-orbit interaction, that dominates. Here we are concerned with the L1$_0$ 
structure of FePt, which is tetragonal, and so a uniaxial MCA effectively models the preference for the magnetization to align along 
the $c$-axis of the crystal.} 

Finite temperature properties are
computed by employing a Langevin approach introduced by Brown\cite{Brown1963}, \new{effectively converting equation (\ref{eq:LLG}) 
into the stochastic LLG equation. In this formalism,} the thermal noise term, $\boldsymbol{\xi}$, behaves as a random Gaussian
variable with mean and variance given by
\begin{align}
  \langle \xi_{i\alpha}(t) \rangle  &= 0, \\
  \langle \xi_{i\alpha}(t) \xi_{j\beta}(t') \rangle &= \frac{2 \gamma \lambda k_\mathrm{B} T}{\mu_i} \delta_{ij} \delta_{\alpha \beta} \delta(t-t').
\end{align}
Here $i,j$ label different atoms, $\alpha, \beta = x,y,z$ are the Cartesian components, $k_\mathrm{B}$ is the Boltzmann constant and
$T$ is the thermodynamic temperature.


In the work of Mryasov \emph{et al.}, {\it ab initio} calculations of L1$_0$ FePt showed that the Pt moment depends on the local
exchange field generated by the Fe atoms. From this Mryasov constructed an `extended spin model' (ESM), where the Pt degrees
of freedom are incorporated into the Fe ones through mediated exchange and anisotropy parameters. This leads to a Hamiltonian
of the form
\begin{equation}
  \mathcal{H}_\text{FePt} = - \sum_{i,j} \tilde{J}_{ij} \mathbf{S}_i \cdot \mathbf{S}_j -\sum_{i,j} d^{(2)}_{ij} S_{iz} S_{jz} - \sum_i d^{(0)} S_{iz}^2\:, \label{eq:ESHam}
\end{equation}
where $\tilde{J}_{ij}$, $d^{(2)}$ and $d^{(0)}$ are the effective exchange, the two-ion anisotropy and the onsite anisotropy, respectively.
Since this model intrinsically takes into account the longitudinal behavior of the Pt moments, Mryasov crucially predicted the relation
$K(T) \propto M(T)^{2.1}$, which within the model originates from the two-ion anisotropy.

\new{The ESM constitutes a valid approach to describe the properties of FePt. Nevertheless, it is characteristic of such material 
and cannot be easily extended to other cases. We propose here a model alternative to the ESM, where the Pt atoms are explicitly 
included in the spin Hamiltonian. This will allow us to reproduce the same thermodynamical properties predicted by Mryasov \emph{et al.} 
and, at the same time, to analyze the interplay between the spins at non-equivalent sites. The dependence of the Pt moments on the 
spins at the Fe sites, however, requires us to relax the constrain of fixed spin length, typical of LLG dynamics.}

\new{A generalization of the LLG equation that includes longitudinal spin fluctuations was already presented by Ma \emph{et al.}\cite{Ma2012b}. 
By considering the spin length to be no longer conserved and by following the analogy of the Langevin equations of motion in molecular dynamics, 
Ma \emph{et al.} constructs an equation of motion that contains both transverse and longitudinal components, which will employ to simulate FePt. 
It reads}
\begin{equation}
  \frac{\partial \mathbf{S}_i}{\partial t} = - \gamma \mathbf{S}_i \times \mathbf{H}_i + \gamma \lambda \mathbf{H}_i + \boldsymbol{\xi}_i\:,\label{eq:long_llg}
\end{equation}
\new{which we term here the generalized spin equation of motion (GSE). It is worth noting that by using the vector triple product identity 
the damping term can be written as 
$ \lambda \mathbf{H}_i = \lambda (\mathbf{S} (\mathbf{S}_i \cdot \mathbf{H}_i) - \mathbf{S}_i \times \mathbf{S}_i \times \mathbf{H}_i) / S^2$. 
This equation of motion can then be seen to contain the conventional Landau-Lifshitz form of damping but also a further longitudinal damping. }

\new{Ma \emph{et al.} connects this equation of motion to an additional energy term.}
The Heisenberg Hamiltonian in Eq.~(\ref{eq:HHam}) is augmented with a longitudinal energy term, $\mathcal{H}_l$, which takes the
shape of a Landau-like Hamiltonian. This contains even powers of the spin length, namely
\begin{equation}
  \mathcal{H}_l = \sum_{\alpha} \sum_{i} A_\alpha |S_i|^2 + B_\alpha |S_i|^4 + C_\alpha |S_i|^6\:,
  \label{eq:LandauHam}
\end{equation}
where \new{$\alpha$ denotes the atomic species (Fe or Pt) and $i$ labels the spins of that species.} $A_\alpha$, $B_\alpha$ and $C_\alpha$ 
are \new{the parameters that determine the shape and energy scale of the longitudinal energy.} Such simple polynomial form can easily be 
implemented into conventional atomistic codes \new{and was calculated by Pan \emph{et al.}\cite{Pan2017} for permalloy.}

For L1$_0$ FePt two sets of parameters
are then required, one for each species. For Fe we adopt here the same parameters \new{calculated by Ma \emph{et al} using first principles 
simulations for bcc Fe. In that work the authors assume that the ferromagnetic ground state is correctly described by the Stoner model.
DFT calculations are then performed in order to estimate the total energy, $E(M)$, for different values of the total magnetization, $M$.
The latter is the electronic analogous of the longitudinal energy, hence the expression in Eq.~(\ref{eq:LandauHam}) can be fitted to $E(M)$.
The resulting parameters are $A_\text{Fe} = \SI{-440.987}{meV}$, $B_\text{Fe} = \SI{150.546}{ meV}$ and $C_\text{Fe} = \SI{50.6794}{meV}$. 
These are strictly computed for the bcc Fe structure and, whilst it is expected that this energy may vary significantly depending on the local atomic
environment, we choose to use the same parameters in lieu of the more detailed first principles calculations. While this is only an approximation
of the true parameter set of FePt, the key detail described here is that there is a parabolic energy minimum located at $|S|=1$.}

For the Pt atoms the energy minimum
is expected to depend linearly on the local exchange field, as predicted by Mryasov \emph{et al.} Therefore the longitudinal energy
must be approximately quadratic with the energy minimum at $|S|=0$. By considering equation (\ref{eq:long_llg}), the equilibrium spin
length is then given by
\begin{align}
  \frac{\partial S_{Pt}}{\partial t} = 0 = & 2 A_\text{Pt} S_{Pt} + 4 B_\text{Pt} S_{Pt}^3 + 6 C_\text{Pt} S_{Pt}^5 \nonumber \\
   & - \sum_j J_{Pt,j} S_j - 2 k_\text{Pt} S_{Pt}, \label{eq:Pt_min}
\end{align}
where we have assumed that all spins are aligned and that those neighboring the Pt sites are at equilibrium, $S_j=1$.
In order to obtain a linear relation between the spin length and the local field we set $B_i = C_i = 0$ and require that the
equilibrium spin length is $S_{Pt} = 1$. Equation (\ref{eq:Pt_min}) then gives
\begin{equation}
  A_\text{Pt} = k_\text{Pt} + \frac{1}{2}\sum_j J_{Pt,j}.
\end{equation}

\begin{figure}
  \centering
  \includegraphics[width=\figwidth]{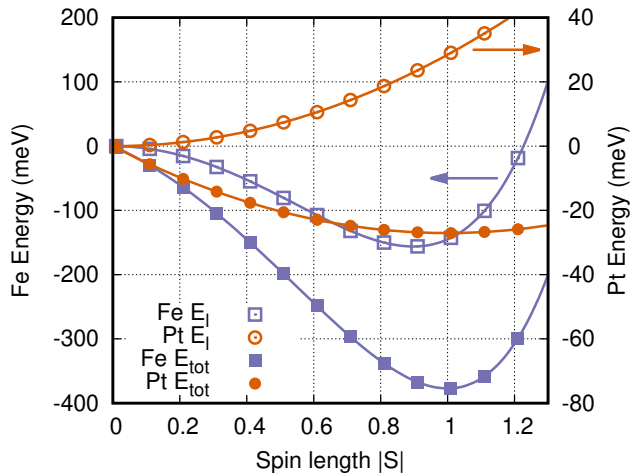}
  \caption{ \label{fig:long_energy}(Color online) Energy as a function of the reduced spin length in FePt. Fe possesses a
  spontaneous moment, which is modeled by having an energy minimum at $|S|=1$, while for Pt the moment is induced
  by the Fe exchange interaction. As such the longitudinal energy is modeled by a quadratic function. The arrows indicate
  the relevant energy scale. The left-hand side scale (right-hand side) is for Fe (Pt).}
\end{figure}

The longitudinal energy for both the Fe and Pt atoms is shown in figure \ref{fig:long_energy}. The open points show the energy
given by equation (\ref{eq:LandauHam}) only while the solid ones show the total Hamiltonian energy with all the neighbors
aligned. For Pt the longitudinal energy has an energy minimum at zero magnetic moment but for the total energy, which includes
the local Fe exchange field, the minimum is at $S=1$, as desired.

From ab-initio calculations we find that the magnetic moment for Fe is $\mu_\text{Fe} = 2.86 \mu_B$ and for Pt (in a ferromagnetic
configuration) is $\mu_\text{Pt} = 0.36 \mu_B$. Mryasov's calculations gives the Pt anisotropy as $k_\text{Pt} = \SI{1.427}{meV}$,
while for the Fe atoms $k_\text{Fe} = \SI{-0.097}{meV}$ as from reference [\onlinecite{Mryasov2004}]. These values give a
macroscopic magnetization of $M_s = \SI{1.072e6}{JT^{-1}m^{-3}}$ and a first-order anisotropy of $K_1 = \SI{7.502e6}{Jm^{-3}}$.
Experimental measurements of the damping parameter vary but it is generally considered to be large due to the high spin-orbit
interaction. Here we use the values found by Becker \emph{et al.}\cite{Becker2014} of $\lambda=0.1$. 
\new{The exchange coupling constants for the Heisenberg Hamiltonian [Eq.~(\ref{eq:HHam})] used here were originally calculated 
by Mryasov et al. in Ref. [\onlinecite{Mryasov2005}] using constrained density functional theory and are summarized in table~\ref{tab:Jij}.} 
The Fe-Pt and Pt-Pt exchange is negligible beyond the nearest neighbor, while the Fe-Fe one is longer ranged. Here we restrict 
the range to the 4th nearest neighbors \new{and have rescaled the parameters so that the total exchange energy is conserved after
truncation}. 
It is worth noting that the in-plane Fe-Fe exchange is stronger than the out-of-plane interactions and the Fe-Pt exchange. 
\new{For the ES model we employ the same mediated exchange parameters calculated by Mryasov et al. and later employed in other 
works investigating properties of FePt\cite{Kazantseva2008PRB,Barker2010,Ellis2015_FePt}. By employing the original isotropic exchange 
interactions computed by Mryasov and the mediated exchange parameters that are derived from them, our calculations using the different 
Hamiltonians have an equivalent exchange energy.}

\begin{table}
\caption{The Heisenberg exchange coupling parameters for the corresponding inter-atomic vector (given in terms of the
conventional unit cell vectors) used for the LLG and GSE models.} \label{tab:Jij}
\begin{ruledtabular}
  \begin{tabular}{l c c c c}  
       & $\vec{a}$& $\vec{b}$ & $\vec{c}$& $J_{ij}$ (meV) \\
    \hline
    Fe-Fe & 1/2 & 1/2 & 0   & 16.356 \\
          & 0   & 0   & 1   & ~~1.762 \\
          & 1   & 0   & 0   & 13.653\\
          & 1/2 & 1/2 & 1   & ~~5.886 \\
    \hline
    Fe-Pt & 1/2 & 0   & 1/2 & ~~6.666 \\
    \hline
    Pt-Pt & 1/2 & 1/2 & 0   & ~~0.177 \\
  \end{tabular}

  \end{ruledtabular}

\end{table}

\begin{figure}
  \centering
  \includegraphics[width=\figwidth]{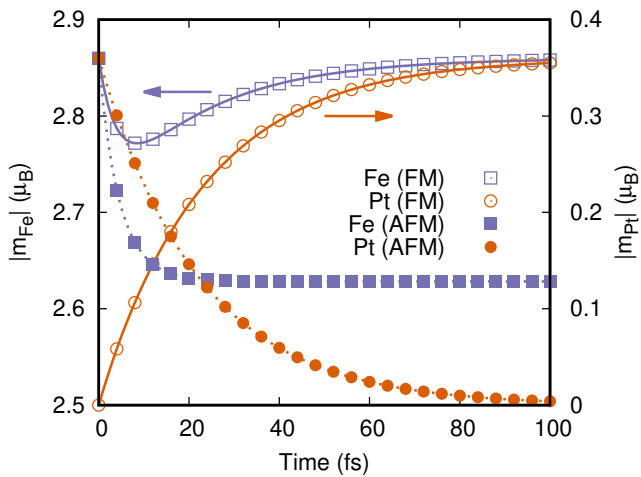}
  \caption{\label{fig:equil_moment} (Color online) The relaxation of the atomic magnetic moments to equilibrium at T=\SI{0}{K} for two cases: (open points - FM) ferromagnetic and (closed points - AFM) anti-ferromagnetic ordering of the Fe atoms. In both cases the Fe moments start fully saturated (i.e $|S| = 1$ whilst the Pt atoms start with $|S|=0$ for the FM case and $|S|=1$ for the AFM case to highlight the relaxation dynamics. The left-hand side scale (right-hand side) is for Fe (Pt).}
\end{figure}

\new{The dynamic evolution of the magnetic moment length ($|m| = |S| \mu_s$) toward equilibrium at $T=$\SI{0}{K} is shown in 
figure~\ref{fig:equil_moment}. We consider two different configurations for the Fe moments: (1) the ferromagnetic ground-state (FM) and 
(2) a quasi-equilibrium anti-ferromagnetic state (AFM), where the Fe moments alternate along the $z$-axis. In both cases the Fe moments 
are initialized to $|S|=1$, while the Pt moments are $|S|=0$ for the FM case and $|S|=1$ for the AFM one. This is done intentionally to highlight 
the dynamics towards the local energy minima. No torque acts on the moments of these initial conditions, since they are collinear, so that only 
longitudinal dynamics takes place. As figure \ref{fig:equil_moment} shows, in the FM case the Pt moments are polarized by the exchange field 
and converges towards $0.36 \mu_B$ ($|S|=1$) while in the AFM case the exchange field cancels and so the Pt moments relax towards the
energy minima of the longitudinal Hamiltonian, which by construction is 0. The Fe moments relax slightly in the AFM configuration due to the 
loss of the exchange from the Pt atoms, but is only a change of approximately 8\%. In the FM configuration there is a short transient associated 
to the Pt moments evolving from 0 to 1.}

In order to compute the finite temperature properties of these models we numerically integrate the LLG and Longitudinal LLG
equations of motion [eqs. (\ref{eq:LLG}) and (\ref{eq:long_llg}), respectively] by using the stochastic Heun scheme\cite{Evans2014}.
Since this method does not conserve the spin length implicitly when integrating the LLG equation the spin is renormalized during
each step while for the GSE model no renormalization step is performed. The time-step used during the simulation is $\Delta t = \SI{0.1}{fs}$,
which is found to be stable for both the LLG and GSE model. In order to confirm our implementation we also compare our static
calculations to that of a Monte-Carlo model. As in Ref.~[\onlinecite{Pan2017}] we chose the phase space measure to be
unitary and for each trial step we displace the spin by an amount taken uniformly from a sphere with a size that is controlled
to attain a 50\% acceptance ratio.

\section{Results and Discussion}

\begin{figure}
  \centering
  \includegraphics[width=\figwidth]{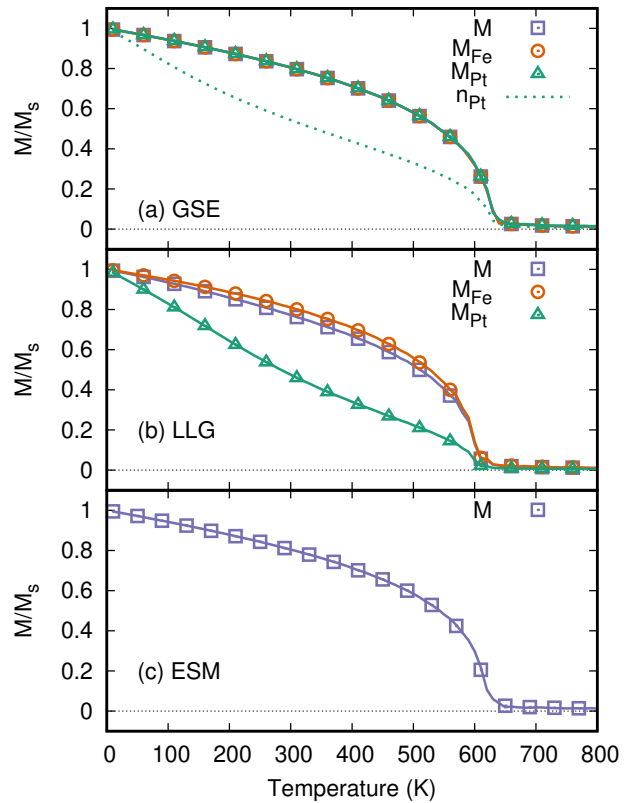}
  \caption{ (Color online) The average magnetization as a function of temperature in L1$_0$ FePt calculated by using (a) the GSE, (b)
  the LLG and (c) the ES model. Lines show results obtained by using the spin dynamics models, while the points
  are for Monte Carlo simulations. Each model gives slightly different Curie temperatures but they all are close
  to the experimental value of \SI{600}{K}. In (a) and (b) the average sub-lattice magnetization of Fe and Pt are plotted
  separately. In (a) the average of the Pt magnetization unit vector, $n_\text{Pt}$, is shown as a dotted line as a comparison
  with (b).}
  \label{fig:MvT}
\end{figure}

We begin by examining the thermodynamic properties of the FePt system. In the following we compare the three models
described in the previous section which, to summarize, are: (ESM) the Fe-only Hamiltonian [Eq.~(\ref{eq:ESHam})] of 
Mryasov \emph{et al.} simulated with the LLG equation; (LLG) the Heisenberg Hamiltonian [Eq.~(\ref{eq:HHam})] including 
Pt moments simulated with the LLG equation (no longitudinal relaxation); (GSE) the Heisenberg-Landau Hamiltonian 
[Eqs. (\ref{eq:HHam}) and (\ref{eq:LandauHam})] with Pt moments simulated by using the \new{generalized spin equation 
of motion given in equation}~(\ref{eq:long_llg}). Figure \ref{fig:MvT} shows the temperature dependence of the magnetization 
calculated, for each model, by using both Monte Carlo and spin dynamics simulations. In all cases Monte Carlo and spin dynamics return essentially
identical magnetization values over the entire temperature range, showing that the equations of motion are integrated correctly.
We find a surprisingly good agreement between the GSE model and ESM with the Curie temperatures, $T_\mathrm{C}$,
found to be \SI{619}{K} and \SI{617}{K}, respectively. In contrast the LLG model returns a slightly lower $T_\mathrm{C}$
(\SI{602}{K}). This is in contrast to what has been observed for {\it bcc} Fe by Ma \emph{et al.}\cite{Ma2012b} and
Pan \emph{et al.}\cite{Pan2017}, in which the $T_\mathrm{C}$ was reduced when including the longitudinal dynamics.
As described by Pan \emph{et al.}, in order to correct for this change in the Curie temperature one must apply a re-scaling
parameter to the exchange constants. Such rescaling factor should be calculated for each material and there is no general
trend in the Curie temperature change upon introducing longitudinal fluctuations.

It is worth noting that in the parameterization of the exchange coupling the Fe-Fe interaction is stronger than the Fe-Pt one
and much longer ranged, as shown in table \ref{tab:Jij}. This results in a fairly rigid
Fe sub-lattice to which the Pt one is coupled to. The average sub-lattice magnetizations are shown for the LLG and GSE
models in panels (a) and (b). In the case of the LLG approach there is significant spin non-collinearity in the Pt sub-lattice
well below $T_\mathrm{C}$, while the total magnetization is dominated by the Fe sites. In the GSE model the Pt magnetization
follows almost identically that of Fe but in this case there are both longitudinal and transverse changes. The dashed line in
panel (a) shows the average of the Pt magnetization unit vector, $n_\mathrm{Pt}(t) = M_\mathrm{Pt}(t)/|M_\mathrm{Pt}(t)|$,
which measures the transverse disorder of the sub-lattice. It shows a similar behavior to the Pt sub-lattice in the LLG model,
which indicates that whilst there is still large transverse disorder due to the weak exchange coupling at the Pt sub-lattice the
magnitude of the local Pt moments increases.

\begin{figure}
  \centering
  \includegraphics[width=\figwidth]{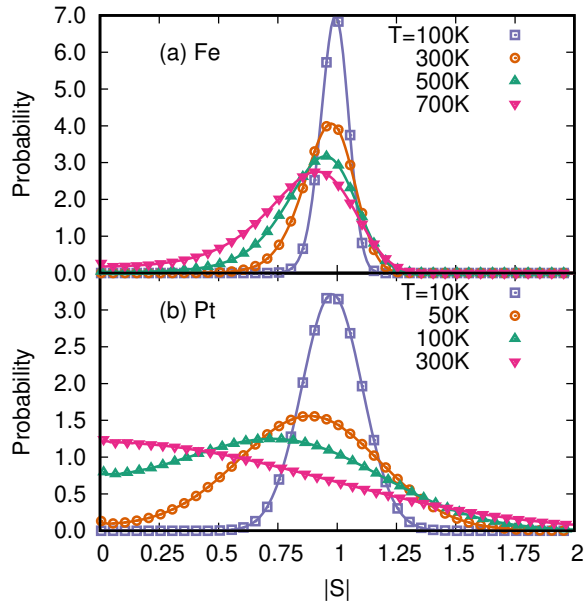}
  \caption{ (Color online) Probability distribution of finding a (a) Fe or (b) a Pt moment with a certain spin length at different temperatures.
  For Fe the probability is peaked at around $|S| = 1$ for all temperatures due to the form of the longitudinal energy function,
  while for Pt is decreases with temperature due to the increasing non-collinearity of the neighboring Fe atoms. }
  \label{fig:Snorm}
\end{figure}

Figure \ref{fig:Snorm} shows the probability distribution of the magnitude of the spin vector at different temperatures. The probability
distribution for Fe is peaked close to $|S| = 1$ for all temperatures even above the Curie temperature, as expected from
the longitudinal energy function. For Pt the peak of the distribution moves to lower spin values with temperature, until about
\SI{200}{K}, where it becomes centered at $|S|=0$. For temperatures above such critical value the distribution widens
continuously, and the upper tail is, in principle, unbound. This effect arises due to the choice of energy function, which does
not constrain the upper bound of the Pt local moment. A more realistic description may include such constraint, which is
ultimately determined by the electron count in the system.

\begin{figure}
  \centering
  \includegraphics[width=\figwidth]{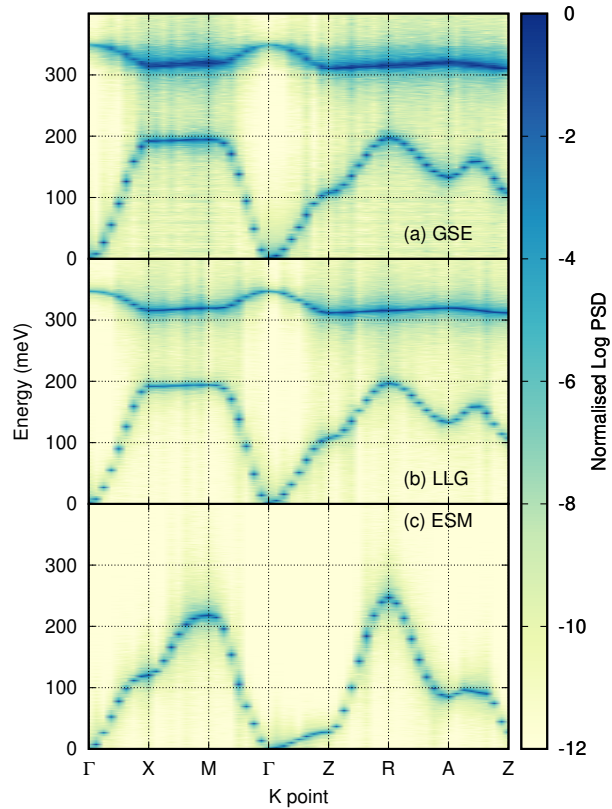}
  \caption{ (Color online) Dynamics structure factor for the (a) GSE, (b) LLG and (c) ES model. The color intensity encodes the logarithm
  of the power spectral density normalized to the peak values along each $k$-vector. Since the GSE and LLG models contain
  two atoms in the primitive cell both acoustic and optical branches are present. Only the acoustic branch is observed for the ESM.}
  \label{fig:spinwaves}
\end{figure}

In order to gain a deeper insight into the properties of the system we next investigate the spin-wave spectra of each model. This
is computed by using the dynamic structure factor (DSF)
\begin{equation}
  \tilde{C}(\mathbf{k}, \omega) = \int e^{-i \omega t} dt \sum_{\mathbf{r},\mathbf{r}'} e^{-i \mathbf{k} \cdot ( \mathbf{r} - \mathbf{r}')} C ( \mathbf{r}-\mathbf{r}', t)\:,
\end{equation}
where $C ( \mathbf{r}-\mathbf{r}', t) = \langle S_x( \mathbf{r}, 0) S_x (\mathbf{r}', t) \rangle$ is the spin-spin correlation function.
Figure \ref{fig:spinwaves} shows the computed DSF for each model at $T=\SI{10}{K}$ along the symmetry lines of the tetragonal
Brillouin zone. The height and width of the peaks depends on the damping. Therefore, in order to obtain a better view of the spin-wave
modes we have reduced the damping parameter to $\lambda = 0.01$ for this figure. Since the ES model incorporates the Pt degrees
of freedom into those of Fe there is only one Fe atom in the primitive unit cell leading to only an acoustic magnon branch.
In contrast the GSE and LLG models include the Pt moments and the second optical branch is observed. Since the exchange
constants in the ES and GSE/LLG models are different the magnon bands do not agree well with each other.

However, while the acoustic branches of the ES and GSE/LLG models do not match at relatively high energy, they show
a rather similar exchange stiffness, $D$, at low $k$. It is then not surprising that the Curie temperatures for the three models are
calculated to be very similar, since these are determined mostly by the low-energy part of the excitation spectrum.
The exchange stiffness, the Curie temperatures and the critical exponents calculated by each model are summarized in Table~\ref{tab:TC} \new{with experimental measurements for comparison. In comparing to the experimental values the models all under-estimate the Curie temperature, due to the DFT computed exchange constants used, and while the ESM is closest to the experimental exchange stiffness value, calculated from the measurements of Antoniak et al.\cite{Antoniak2010}, all three models are within the error range}
Interestingly, the exchange stiffness for the ESM is slightly lower than that of the LLG value, despite its Curie
temperature being slightly higher. Such small anomaly is then explained with the contribution of the high-$k$ spin-wave
excitations to the $T_\mathrm{C}$. Key differences between the ES and the GSE/LLG models are found at the
Z and X high-symmetry points in the Brillouin zone, with GSE/LLG returning always a significantly larger magnon
energy. As a consequence the energy dispersion along the $\Gamma-Z$ is much more pronounced for the GSE/LLG
models, while it is rather flat for the ESM. These differences arise from the exchange interactions, which in the ESM
are altered through the mediation of the Pt lattice. In relation to the GSE model the bands are of similar character to
the LLG one with the exception of an increased line-width close to the edge of the Brillouin zone, particularly in the
optical branch.

\begin{table}
\caption{\label{tab:TC} The Curie temperature, $T_\mathrm{C}$, the critical exponent, $\beta$, and the exchange stiffness, $D$,
calculated from the GSE, LLG and ES models.}
\begin{ruledtabular}
  \renewcommand{\arraystretch}{1.25}
\begin{tabular}{l|ccc}
   &$T_C$ (K)& $\beta$ & D (\new{\si{meV\AA^{2}}})\\
  \hline
  GSE & 619.1 $\pm$ 0.3 & 0.329 $\pm$ 0.001 & 304.74          $\pm$ 0.02\\
  LLG & 602.6 $\pm$ 0.1 & 0.370 $\pm$ 0.001 & 301.64 $\pm$ 0.03\\
  ESM & 617.4 $\pm$ 0.9 & 0.326 $\pm$ 0.001 & 275.85 $\pm$ 0.13\\
  Exp. & 750\footnote{From Ref. \onlinecite{Weller2000}.} &       & 257 $\pm$ 86\footnote{Derived from measurements of the exchange stiffness in Ref. \onlinecite{Antoniak2010}}. \\
\end{tabular}
\end{ruledtabular}

\end{table}

\begin{figure}
  \centering
  \includegraphics[width=\figwidth]{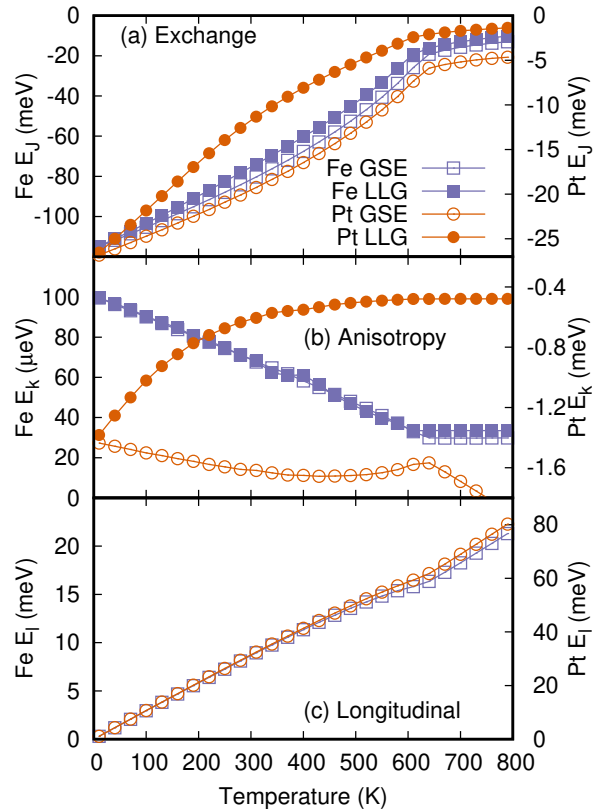}
  \caption{ (Color online) The temperature dependence of the average Hamiltonian energies in FePt separated into the three
  main contributions: (a) exchange ($E_J$), (b) anisotropy ($E_k$) and (c) longitudinal ($E_l$) for each element. Solid (open)
  symbols correspond to the LLG (GSE) model. The Fe energies are given on the left-hand side axis and the Pt ones are on the
  right-hand side. In (c) $E_l$ is given relative to the $T=$~\SI{0}{K} values.
  The exchange energy shows a consistent decrease (in magnitude) with temperature. For Fe the GSE and LLG models are
  similar while for Pt there is a change in the power scaling from $<1$ for GSE to $>1$ for LLG. Likewise the anisotropy energy for
  Fe behaves similarly for the two models whilst that of Pt decreases with $T$ for GSE and increases for LLG. The longitudinal
  energy in the GSE model increases with $T$ for both Fe and Pt but on a different scale.}
  \label{fig:FePt_EvT}
\end{figure}

We now turn to examine the average internal energies as a function of temperature, which are shown in
figure~\ref{fig:FePt_EvT}. Let us discuss first the average exchange energy, shown in panel (a). For the
LLG model (solid symbols) the Pt exchange energy drops as a power law with an exponent $>1$, while
for the GSE one (open symbols) the exponent is more similar to that of the Fe exchange energy, which is
$<1$. The Fe exchange energy of both models has an almost identical temperature dependence with the
exception of a small shift in the Curie temperature, as noted already in figure \ref{fig:MvT}. Since the exchange
energy is a measure of the non-collinearity of the system, we ascribe the difference in the behavior of the
Pt contribution to the fact that the magnetization of Pt follows that of Fe in a much closer way for the GSE model
than for the LLG one.

The anisotropy energy [panel (b)] shows again that the two models behave quite similarly at the Fe sub-lattice,
but they are markedly different at the Pt one. The Pt anisotropy energy decreases rapidly with temperature in
the LLG model, while in the GSE model it slightly increases, a behavior that we attribute to two factors. Firstly, as
seen for the exchange energy, the Pt moments are more aligned with that of the Fe within the LLG model. Secondly,
the spin length is increased in the GSE model. This second factor can be observed above the Curie temperature.
In fact, in the paramagnetic state there is a uniform angular distribution of the Pt spins but, as seen from the probability
distributions [figure~\ref{fig:Snorm}(b)], there are also spins with a significantly large length, a factor that affects
the anisotropy energy.

Finally, the longitudinal energy [panel (c)] shows an increase with temperature for both the Fe and Pt moments
but on a much larger scale in Pt. Again, we attribute this behavior to the widening distribution of the Pt spin lengths
leading to an occupation of higher energy states. This additional energy component provides another degree of
freedom for the internal energy to be distributed amongst. This may explain the changes in the other energy components.

%
%

%

\begin{figure}
  \centering
  \includegraphics[width=\figwidth]{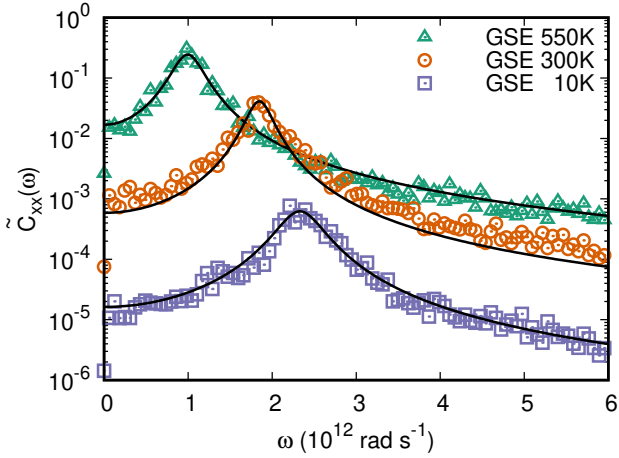}
  \caption{ (Color online) Fourier transform of the $m_x$ correlation function at 10~K, 300~K and 550~K. The solid lines show
  a fit using equation~(\ref{eq:FMR}). With increasing temperature the resonance field drops and so does the magnitude
  of the background white noise.}
  \label{fig:FePt_corr}
\end{figure}

When compared to the anisotropy calculated in figure \ref{fig:FePt_EvT} the macroscopic one contains also entropic
contributions. Such macroscopic anisotropy can then be computed in a ferromagnetic resonance (FMR) type simulation.
In fact, both the anisotropy field and the macroscopic damping coefficient can be determined from the FMR line shape,
thus allowing us to understand the effect of the induced Pt moments on the dynamic response of the system. The resonant
FMR peak can be extracted from the dynamics via the magnetization correlations function
\begin{equation}
  \tilde{C}_{xx}(\omega) = \int e^{-i \omega t} dt \langle m_x(0) m_x(t) \rangle\:,
\end{equation}
where $m_x(t)$ is the $x$ component of the magnetization at time $t$. This is related to the dynamic susceptibility through
the fluctuation-dissipation theorem\cite{Kubo1966}. By using linear response theory\cite{Butera2006} a form for the line shape is
found to be
\begin{equation}
  \tilde{C}_{xx}(\omega) = \frac{2 k_B T \gamma \alpha}{1+\alpha^2} \left( \frac{ \omega_0^2(1+\alpha^2) + \omega^2}{\Omega^4 + (2\alpha \omega_0 \omega)^2} \right),\label{eq:FMR}
\end{equation}
where $\Omega^2 = \omega_0^2(1+\alpha^2) - \omega^2$, $\omega_0 = \gamma H_z / (1+\alpha^2)$ is the resonance
frequency and  $\alpha$ is the Gilbert damping coefficient describing the relaxation of the magnetization vector. $H_z$ is
the field in the $z$-direction, which in this case is given by the anisotropy field.

The dynamic correlation function computed using the GSE model is shown in figure \ref{fig:FePt_corr} at \SI{10}{K}, \new{\SI{300}{K}} and \SI{550}{K}.
This is determined by computing a \SI{2}{ns}-long time-series with a sample rate of \SI{10}{THz}. The power spectral density is
then calculated by using Welch's method of separating the time-series into blocks of 2048 samples. At all temperatures a
clear resonance peak is observed with a line-width related to the Gilbert damping of the system. As the temperature increases
the resonance field drops due to a reduction in the macroscopic anisotropy. At high frequency, away from the resonance peak,
the spectral profile is flat, in agreement with the white noise approximation of stochastic micromagnetic models\cite{Evans2012}.
The lines in figure~\ref{fig:FePt_corr} show a fit to the data obtained by using equation~(\ref{eq:FMR}) with resonance frequency,
Gilbert damping and the pre-factor taken as fitting parameters. As shown in figure~\ref{fig:FePt_corr} the function fits the data
well but due to the large damping used ($\lambda=0.1$) the signal-to-noise ratio is poor, giving a large data scatter for the Gilbert damping.
\new{While equation~(\ref{eq:FMR}) is derived within a linear response approach we find that it fits well close to $T_\mathrm{C}$. 
This appears to be due to the large anisotropy keeping the magnetization aligned along the $z$-axis even close to $T_\mathrm{C}$, 
with the $x$ and $y$ components of the magnetization being small despite the thermal fluctuations.}

\begin{figure}
  \centering
  \includegraphics[width=\figwidth]{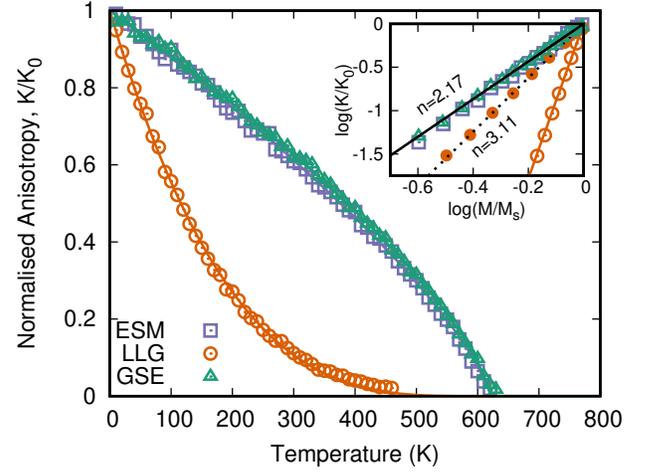}
  \caption{ (Color online) Anisotropy field as a function of temperature calculated from the FMR for each model. The ESM and GSE
  models drop slowly until reaching the Curie temperature, while the LLG model shows a fast decay with temperature even well
  below T$_\mathrm{C}$. The inset shows the scaling of the anisotropy with the total magnetization, showing a power of $n=2.17$
  for the GSE and ES models, while for the LLG model we find $n=8.90$. For the LLG model we also present the case where the
  anisotropy is scaled with the Pt sub-lattice magnetization (solid orange circles). In this case the scaling follows a $n=3.1$ power
  law, closer to the $n=3$ expected for pure uniaxial anisotropy.}
  \label{fig:FePt_Hk}
\end{figure}

\begin{figure}
  \centering
  \includegraphics[width=\figwidth]{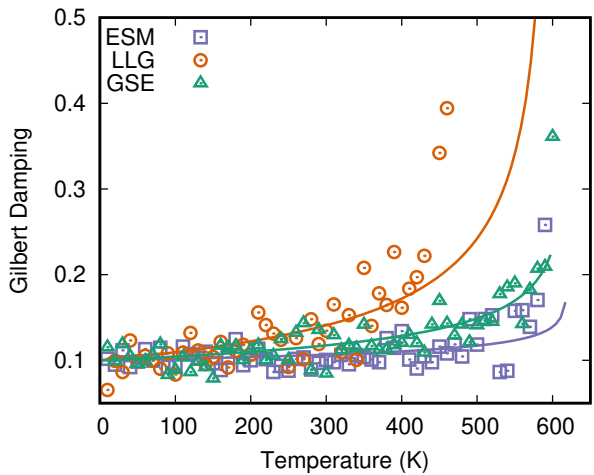}
  \caption{ (Color online) Temperature dependence of the Gilbert damping extracted from fitting the FMR line-shape. At low
  temperature the Gilbert damping is close to the atomistic damping value ($\lambda=0.1$), while it diverges as we approach the Curie temperature.}
  \label{fig:FePt_FMR_damping}
\end{figure}

From fitting the lineshape of the correlation function the anisotropy field has been extracted. Figure \ref{fig:FePt_Hk}
shows the temperature dependence of the anisotropy field in FePt calculated with our three different models. The
GSE model and ESM exhibit very similar behavior, which is in sharp contrast with that predicted by LLG. The inset
of figure \ref{fig:FePt_Hk} shows the power law scaling of the macroscopic anisotropy with the total magnetization.
Both GSE and ESM return a power scaling with an exponent of $\approx 2.17$, which is close to the experimentally
measured value\cite{Okamoto2002} of 2.1.
\new{As shown by Mryasov \emph{et al.}\cite{Mryasov2005}, the ESM finds the $n=2.1$ exponent because of the two-ion 
anisotropy mediated by the Pt moments. Remarkably, the GSE model finds a similar exponent despite the only the uniaxial 
anisotropy contribution to the Hamiltonian. The longitudinal fluctuations are therefore important in providing this two-ion-like 
anisotropy naturally and without a complicated re-parameterization of the Hamiltonian. Additional parameters 
are required for the Landau Hamiltonian, but since the parameters for Pt are determined by the exchange interactions this 
approach appears robust.} The LLG model with fixed spin length, in contrast, gives us the vastly different
power of $n=8.90$, which is not only in disagreement with experiments, but also disagrees with the theory
for pure uniaxial anisotropy\cite{Callen1966}, which predicts $n=3$. However, when we consider the scaling with respect
to the Pt sub-lattice magnetization (solid circles) we find that this follows a $n=3.1$ power-law.

Finally, figure~\ref{fig:FePt_FMR_damping} shows the macroscopic Gilbert damping extracted from the FMR simulations.
At low temperatures the Gilbert damping for all the three models remains close to the value of the atomistic damping, $\lambda=0.1$.
However, as the temperature approaches the Curie point the Gilbert damping diverges. The approach to such divergence is different
for the three models, with the LLG one showing a clearly more rapid increase of $\alpha$ with temperature. In fact, in
this case the anisotropy field disappears above $\approx \SI{450}{K}$ so that the damping cannot be extracted closer to
$T_\mathrm{C}$. When comparing GSE to  ESM the calculated Gilbert damping are quite similar, although for GSE $\alpha$
seems to remain constantly above than that of ESM at any temperature. This can be understood from the fact that the GSE
model presents additional relaxation channels due to the explicit presence of the Pt moments. Our analysis thus shows
the advantage of the GSE model over the alternatives as it describes the Pt dynamics out of the ground state. This appears to
be important. Recent ultrafast experiments have, in fact, directly observed differing timescales for the Pt and Fe
magnetic moments\cite{Yamamoto2018a}.

\section{Conclusion}
In conclusion, the role of the induced Pt magnetic moments in ordered L1$_0$ FePt has been studied by using a model
of longitudinal fluctuations. This has been constructed to allow the Pt magnetic moment to vary linearly with the local exchange
field as predicted by previous {\it ab initio} calculations. The longitudinal fluctuation model has been compared to the existing
extended spin Hamiltonian approach of Mryasov \emph{et al.}, showing similar results concerning the FePt static properties.
An analysis of the spin magnitude histogram as a function of temperature shows that the Fe moment remains constant, while
for Pt above 200~K the moment distribution is centered around zero, but it presents a long tail of large moments. The dynamic
structure factor shows that the magnon branches are significantly different and the inclusion of the longitudinal fluctuations
leads to a broadening of the magnon modes at the edge of the Brillouin zone. This hints that the dynamical properties must
be different for the different models, as confirmed by our FMR analysis. In particular we find that the anisotropy field exhibits the
experimentally observed $K(T) \propto M(T)^{2.1}$ power scaling without any need to consider a two-ion anisotropy used in the
extended spin Hamiltonian model \new{as it is naturally included in the longitudinal dynamics of the Pt moments driven by the local field of the neighboring Fe moments}. This is a critical advantage of our Hamiltonian, namely the key thermodynamic and dynamical properties
of the material are correctly observed without performing the complex re-parameterization done by Mryasov et al. In contrast, we describe all relevant degrees
of freedom on the same footing, a fact that may allow us to unlock and understand out-of-equilibrium phenomena at fast
timescales.

\section{Acknowledgements}
This work has been supported by the Science Foundation Ireland Principal Investigator award (Grants No. 14/IA/2624 and No. 16/US-C2C/3287) and TCHPC (Research IT, Trinity College Dublin). The authors wish to acknowledge the DJEI/DES/SFI/HEA Irish Centre for High-End Computing (ICHEC) for the provision of computational facilities and support.

\bibliography{library.bib}

\begin{thebibliography}{27}%
\makeatletter
\providecommand \@ifxundefined [1]{%
 \@ifx{#1\undefined}
}%
\providecommand \@ifnum [1]{%
 \ifnum #1\expandafter \@firstoftwo
 \else \expandafter \@secondoftwo
 \fi
}%
\providecommand \@ifx [1]{%
 \ifx #1\expandafter \@firstoftwo
 \else \expandafter \@secondoftwo
 \fi
}%
\providecommand \natexlab [1]{#1}%
\providecommand \enquote  [1]{``#1''}%
\providecommand \bibnamefont  [1]{#1}%
\providecommand \bibfnamefont [1]{#1}%
\providecommand \citenamefont [1]{#1}%
\providecommand \href@noop [0]{\@secondoftwo}%
\providecommand \href [0]{\begingroup \@sanitize@url \@href}%
\providecommand \@href[1]{\@@startlink{#1}\@@href}%
\providecommand \@@href[1]{\endgroup#1\@@endlink}%
\providecommand \@sanitize@url [0]{\catcode `\\12\catcode `\$12\catcode
  `\&12\catcode `\#12\catcode `\^12\catcode `\_12\catcode `\%12\relax}%
\providecommand \@@startlink[1]{}%
\providecommand \@@endlink[0]{}%
\providecommand \url  [0]{\begingroup\@sanitize@url \@url }%
\providecommand \@url [1]{\endgroup\@href {#1}{\urlprefix }}%
\providecommand \urlprefix  [0]{URL }%
\providecommand \Eprint [0]{\href }%
\providecommand \doibase [0]{http://dx.doi.org/}%
\providecommand \selectlanguage [0]{\@gobble}%
\providecommand \bibinfo  [0]{\@secondoftwo}%
\providecommand \bibfield  [0]{\@secondoftwo}%
\providecommand \translation [1]{[#1]}%
\providecommand \BibitemOpen [0]{}%
\providecommand \bibitemStop [0]{}%
\providecommand \bibitemNoStop [0]{.\EOS\space}%
\providecommand \EOS [0]{\spacefactor3000\relax}%
\providecommand \BibitemShut  [1]{\csname bibitem#1\endcsname}%
\let\auto@bib@innerbib\@empty
\bibitem [{\citenamefont {Weller}\ \emph {et~al.}(2000)\citenamefont {Weller},
  \citenamefont {Moser}, \citenamefont {Folks},\ and\ \citenamefont
  {Best}}]{Weller2000}%
  \BibitemOpen
  \bibfield  {author} {\bibinfo {author} {\bibfnamefont {D.}~\bibnamefont
  {Weller}}, \bibinfo {author} {\bibfnamefont {A.}~\bibnamefont {Moser}},
  \bibinfo {author} {\bibfnamefont {L.}~\bibnamefont {Folks}}, \ and\ \bibinfo
  {author} {\bibfnamefont {M.}~\bibnamefont {Best}},\ }\href
  {http://ieeexplore.ieee.org/xpls/abs{\_}all.jsp?arnumber=824418} {\bibfield
  {journal} {\bibinfo  {journal} {IEEE Trans. Magn.}\ }\textbf {\bibinfo
  {volume} {36}},\ \bibinfo {pages} {10} (\bibinfo {year} {2000})}\BibitemShut
  {NoStop}%
\bibitem [{\citenamefont {Okamoto}\ \emph {et~al.}(2002)\citenamefont
  {Okamoto}, \citenamefont {Kikuchi}, \citenamefont {Kitakami}, \citenamefont
  {Miyazaki}, \citenamefont {Shimada},\ and\ \citenamefont
  {Fukamichi}}]{Okamoto2002}%
  \BibitemOpen
  \bibfield  {author} {\bibinfo {author} {\bibfnamefont {S.}~\bibnamefont
  {Okamoto}}, \bibinfo {author} {\bibfnamefont {N.}~\bibnamefont {Kikuchi}},
  \bibinfo {author} {\bibfnamefont {O.}~\bibnamefont {Kitakami}}, \bibinfo
  {author} {\bibfnamefont {T.}~\bibnamefont {Miyazaki}}, \bibinfo {author}
  {\bibfnamefont {Y.}~\bibnamefont {Shimada}}, \ and\ \bibinfo {author}
  {\bibfnamefont {K.}~\bibnamefont {Fukamichi}},\ }\href {\doibase
  10.1103/PhysRevB.66.024413} {\bibfield  {journal} {\bibinfo  {journal} {Phys.
  Rev. B}\ }\textbf {\bibinfo {volume} {66}},\ \bibinfo {pages} {024413}
  (\bibinfo {year} {2002})}\BibitemShut {NoStop}%
\bibitem [{\citenamefont {Zener}(1954)}]{Zener1954}%
  \BibitemOpen
  \bibfield  {author} {\bibinfo {author} {\bibfnamefont {C.}~\bibnamefont
  {Zener}},\ }\href {\doibase 10.1103/PhysRev.96.1335} {\bibfield  {journal}
  {\bibinfo  {journal} {Phys. Rev.}\ }\textbf {\bibinfo {volume} {96}},\
  \bibinfo {pages} {1335} (\bibinfo {year} {1954})}\BibitemShut {NoStop}%
\bibitem [{\citenamefont {Callen}\ and\ \citenamefont
  {Callen}(1966)}]{Callen1966}%
  \BibitemOpen
  \bibfield  {author} {\bibinfo {author} {\bibfnamefont {H.}~\bibnamefont
  {Callen}}\ and\ \bibinfo {author} {\bibfnamefont {E.}~\bibnamefont
  {Callen}},\ }\href {\doibase 10.1016/0022-3697(66)90012-6} {\bibfield
  {journal} {\bibinfo  {journal} {J. Phys. Chem. Solids}\ }\textbf {\bibinfo
  {volume} {27}},\ \bibinfo {pages} {1271} (\bibinfo {year}
  {1966})}\BibitemShut {NoStop}%
\bibitem [{\citenamefont {Skomski}\ \emph {et~al.}(2003)\citenamefont
  {Skomski}, \citenamefont {Kashyap},\ and\ \citenamefont
  {Sellmyer}}]{Skomski2003}%
  \BibitemOpen
  \bibfield  {author} {\bibinfo {author} {\bibfnamefont {R.}~\bibnamefont
  {Skomski}}, \bibinfo {author} {\bibfnamefont {A.}~\bibnamefont {Kashyap}}, \
  and\ \bibinfo {author} {\bibfnamefont {D.}~\bibnamefont {Sellmyer}},\ }\href
  {\doibase 10.1109/TMAG.2003.815746} {\bibfield  {journal} {\bibinfo
  {journal} {IEEE Trans. Magn.}\ }\textbf {\bibinfo {volume} {39}},\ \bibinfo
  {pages} {2917} (\bibinfo {year} {2003})}\BibitemShut {NoStop}%
\bibitem [{\citenamefont {Daalderop}\ \emph {et~al.}(1991)\citenamefont
  {Daalderop}, \citenamefont {Kelly},\ and\ \citenamefont
  {Schuurmans}}]{Daalderop1991}%
  \BibitemOpen
  \bibfield  {author} {\bibinfo {author} {\bibfnamefont {G.~H.~O.}\
  \bibnamefont {Daalderop}}, \bibinfo {author} {\bibfnamefont {P.~J.}\
  \bibnamefont {Kelly}}, \ and\ \bibinfo {author} {\bibfnamefont {M.~F.~H.}\
  \bibnamefont {Schuurmans}},\ }\href {\doibase 10.1103/PhysRevB.44.12054}
  {\bibfield  {journal} {\bibinfo  {journal} {Phys. Rev. B}\ }\textbf {\bibinfo
  {volume} {44}},\ \bibinfo {pages} {12054} (\bibinfo {year}
  {1991})}\BibitemShut {NoStop}%
\bibitem [{\citenamefont {Mryasov}\ \emph {et~al.}(2005)\citenamefont
  {Mryasov}, \citenamefont {Nowak}, \citenamefont {Guslienko},\ and\
  \citenamefont {Chantrell}}]{Mryasov2005}%
  \BibitemOpen
  \bibfield  {author} {\bibinfo {author} {\bibfnamefont {O.~N.}\ \bibnamefont
  {Mryasov}}, \bibinfo {author} {\bibfnamefont {U.}~\bibnamefont {Nowak}},
  \bibinfo {author} {\bibfnamefont {K.~Y.}\ \bibnamefont {Guslienko}}, \ and\
  \bibinfo {author} {\bibfnamefont {R.~W.}\ \bibnamefont {Chantrell}},\ }\href
  {\doibase 10.1209/epl/i2004-10404-2} {\bibfield  {journal} {\bibinfo
  {journal} {Europhys. Lett.}\ }\textbf {\bibinfo {volume} {69}},\ \bibinfo
  {pages} {805} (\bibinfo {year} {2005})}\BibitemShut {NoStop}%
\bibitem [{\citenamefont {Hinzke}\ \emph {et~al.}(2008)\citenamefont {Hinzke},
  \citenamefont {Kazantseva}, \citenamefont {Nowak}, \citenamefont {Mryasov},
  \citenamefont {Asselin},\ and\ \citenamefont {Chantrell}}]{Hinzke2008}%
  \BibitemOpen
  \bibfield  {author} {\bibinfo {author} {\bibfnamefont {D.}~\bibnamefont
  {Hinzke}}, \bibinfo {author} {\bibfnamefont {N.}~\bibnamefont {Kazantseva}},
  \bibinfo {author} {\bibfnamefont {U.}~\bibnamefont {Nowak}}, \bibinfo
  {author} {\bibfnamefont {O.~N.}\ \bibnamefont {Mryasov}}, \bibinfo {author}
  {\bibfnamefont {P.}~\bibnamefont {Asselin}}, \ and\ \bibinfo {author}
  {\bibfnamefont {R.~W.}\ \bibnamefont {Chantrell}},\ }\href {\doibase
  10.1103/PhysRevB.77.094407} {\bibfield  {journal} {\bibinfo  {journal} {Phys.
  Rev. B}\ }\textbf {\bibinfo {volume} {77}},\ \bibinfo {pages} {094407}
  (\bibinfo {year} {2008})}\BibitemShut {NoStop}%
\bibitem [{\citenamefont {Kazantseva}\ \emph {et~al.}(2008)\citenamefont
  {Kazantseva}, \citenamefont {Hinzke}, \citenamefont {Nowak}, \citenamefont
  {Chantrell}, \citenamefont {Atxitia},\ and\ \citenamefont
  {Chubykalo-Fesenko}}]{Kazantseva2008PRB}%
  \BibitemOpen
  \bibfield  {author} {\bibinfo {author} {\bibfnamefont {N.}~\bibnamefont
  {Kazantseva}}, \bibinfo {author} {\bibfnamefont {D.}~\bibnamefont {Hinzke}},
  \bibinfo {author} {\bibfnamefont {U.}~\bibnamefont {Nowak}}, \bibinfo
  {author} {\bibfnamefont {R.~W.}\ \bibnamefont {Chantrell}}, \bibinfo {author}
  {\bibfnamefont {U.}~\bibnamefont {Atxitia}}, \ and\ \bibinfo {author}
  {\bibfnamefont {O.}~\bibnamefont {Chubykalo-Fesenko}},\ }\href {\doibase
  10.1103/PhysRevB.77.184428} {\bibfield  {journal} {\bibinfo  {journal} {Phys.
  Rev. B}\ }\textbf {\bibinfo {volume} {77}},\ \bibinfo {pages} {184428}
  (\bibinfo {year} {2008})}\BibitemShut {NoStop}%
\bibitem [{\citenamefont {Ellis}\ and\ \citenamefont
  {Chantrell}(2015)}]{Ellis2015_FePt}%
  \BibitemOpen
  \bibfield  {author} {\bibinfo {author} {\bibfnamefont {M.~O.~A.}\
  \bibnamefont {Ellis}}\ and\ \bibinfo {author} {\bibfnamefont {R.~W.}\
  \bibnamefont {Chantrell}},\ }\href {\doibase
  http://dx.doi.org/10.1063/1.4919051} {\bibfield  {journal} {\bibinfo
  {journal} {Appl. Phys. Lett.}\ }\textbf {\bibinfo {volume} {106}},\ \bibinfo
  {pages} {162407} (\bibinfo {year} {2015})}\BibitemShut {NoStop}%
\bibitem [{\citenamefont {Ellis}\ \emph {et~al.}(2016)\citenamefont {Ellis},
  \citenamefont {Fullerton},\ and\ \citenamefont {Chantrell}}]{Ellis2016}%
  \BibitemOpen
  \bibfield  {author} {\bibinfo {author} {\bibfnamefont {M.~O.~A.}\
  \bibnamefont {Ellis}}, \bibinfo {author} {\bibfnamefont {E.~E.}\ \bibnamefont
  {Fullerton}}, \ and\ \bibinfo {author} {\bibfnamefont {R.~W.}\ \bibnamefont
  {Chantrell}},\ }\href {\doibase 10.1038/srep30522} {\bibfield  {journal}
  {\bibinfo  {journal} {Sci. Rep.}\ }\textbf {\bibinfo {volume} {6}},\ \bibinfo
  {pages} {30522} (\bibinfo {year} {2016})},\ \Eprint
  {http://arxiv.org/abs/1605.00835} {arXiv:1605.00835} \BibitemShut {NoStop}%
\bibitem [{\citenamefont {Galante}\ \emph {et~al.}(2019)\citenamefont
  {Galante}, \citenamefont {Ellis},\ and\ \citenamefont
  {Sanvito}}]{Galante2019}%
  \BibitemOpen
  \bibfield  {author} {\bibinfo {author} {\bibfnamefont {M.}~\bibnamefont
  {Galante}}, \bibinfo {author} {\bibfnamefont {M.~O.~A.}\ \bibnamefont
  {Ellis}}, \ and\ \bibinfo {author} {\bibfnamefont {S.}~\bibnamefont
  {Sanvito}},\ }\href {\doibase 10.1103/PhysRevB.99.014401} {\bibfield
  {journal} {\bibinfo  {journal} {Phys. Rev. B}\ }\textbf {\bibinfo {volume}
  {99}},\ \bibinfo {pages} {014401} (\bibinfo {year} {2019})}\BibitemShut
  {NoStop}%
\bibitem [{\citenamefont {Yamamoto}\ \emph {et~al.}(2019)\citenamefont
  {Yamamoto}, \citenamefont {Kubota}, \citenamefont {Suzuki}, \citenamefont
  {Hirata}, \citenamefont {Carva}, \citenamefont {Berritta}, \citenamefont
  {Takubo}, \citenamefont {Uemura}, \citenamefont {Fukaya}, \citenamefont
  {Tanaka}, \citenamefont {Nishimura}, \citenamefont {Ohkochi}, \citenamefont
  {Katayama}, \citenamefont {Togashi}, \citenamefont {Tamasaku}, \citenamefont
  {Yabashi}, \citenamefont {Tanaka}, \citenamefont {Seki}, \citenamefont
  {Takanashi}, \citenamefont {Oppeneer},\ and\ \citenamefont
  {Wadati}}]{Yamamoto2018a}%
  \BibitemOpen
  \bibfield  {author} {\bibinfo {author} {\bibfnamefont {K.}~\bibnamefont
  {Yamamoto}}, \bibinfo {author} {\bibfnamefont {Y.}~\bibnamefont {Kubota}},
  \bibinfo {author} {\bibfnamefont {M.}~\bibnamefont {Suzuki}}, \bibinfo
  {author} {\bibfnamefont {Y.}~\bibnamefont {Hirata}}, \bibinfo {author}
  {\bibfnamefont {K.}~\bibnamefont {Carva}}, \bibinfo {author} {\bibfnamefont
  {M.}~\bibnamefont {Berritta}}, \bibinfo {author} {\bibfnamefont
  {K.}~\bibnamefont {Takubo}}, \bibinfo {author} {\bibfnamefont
  {Y.}~\bibnamefont {Uemura}}, \bibinfo {author} {\bibfnamefont
  {R.}~\bibnamefont {Fukaya}}, \bibinfo {author} {\bibfnamefont
  {K.}~\bibnamefont {Tanaka}}, \bibinfo {author} {\bibfnamefont
  {W.}~\bibnamefont {Nishimura}}, \bibinfo {author} {\bibfnamefont
  {T.}~\bibnamefont {Ohkochi}}, \bibinfo {author} {\bibfnamefont
  {T.}~\bibnamefont {Katayama}}, \bibinfo {author} {\bibfnamefont
  {T.}~\bibnamefont {Togashi}}, \bibinfo {author} {\bibfnamefont
  {K.}~\bibnamefont {Tamasaku}}, \bibinfo {author} {\bibfnamefont
  {M.}~\bibnamefont {Yabashi}}, \bibinfo {author} {\bibfnamefont
  {Y.}~\bibnamefont {Tanaka}}, \bibinfo {author} {\bibfnamefont
  {T.}~\bibnamefont {Seki}}, \bibinfo {author} {\bibfnamefont {K.}~\bibnamefont
  {Takanashi}}, \bibinfo {author} {\bibfnamefont {P.~M.}\ \bibnamefont
  {Oppeneer}}, \ and\ \bibinfo {author} {\bibfnamefont {H.}~\bibnamefont
  {Wadati}},\ }\href {\doibase 10.1088/1367-2630/ab5ac2} {\bibfield  {journal}
  {\bibinfo  {journal} {New J. Phys.}\ }\textbf {\bibinfo {volume} {21}},\
  \bibinfo {pages} {123010} (\bibinfo {year} {2019})}\BibitemShut {NoStop}%
\bibitem [{\citenamefont {Ma}\ and\ \citenamefont {Dudarev}(2012)}]{Ma2012b}%
  \BibitemOpen
  \bibfield  {author} {\bibinfo {author} {\bibfnamefont {P.-W.}\ \bibnamefont
  {Ma}}\ and\ \bibinfo {author} {\bibfnamefont {S.~L.}\ \bibnamefont
  {Dudarev}},\ }\href {\doibase 10.1103/PhysRevB.86.054416} {\bibfield
  {journal} {\bibinfo  {journal} {Phys. Rev. B}\ }\textbf {\bibinfo {volume}
  {86}},\ \bibinfo {pages} {054416} (\bibinfo {year} {2012})}\BibitemShut
  {NoStop}%
\bibitem [{\citenamefont {Landau}\ and\ \citenamefont
  {Lifshitz}(1935)}]{Landau1935}%
  \BibitemOpen
  \bibfield  {author} {\bibinfo {author} {\bibfnamefont {L.~D.}\ \bibnamefont
  {Landau}}\ and\ \bibinfo {author} {\bibfnamefont {E.~M.}\ \bibnamefont
  {Lifshitz}},\ }\href@noop {} {\bibfield  {journal} {\bibinfo  {journal}
  {Phys. Z. Sowietunion}\ }\textbf {\bibinfo {volume} {8}},\ \bibinfo {pages}
  {153} (\bibinfo {year} {1935})}\BibitemShut {NoStop}%
\bibitem [{\citenamefont {Gilbert}(2004)}]{Gilbert2004}%
  \BibitemOpen
  \bibfield  {author} {\bibinfo {author} {\bibfnamefont {T.}~\bibnamefont
  {Gilbert}},\ }\href {\doibase 10.1109/TMAG.2004.836740} {\bibfield  {journal}
  {\bibinfo  {journal} {IEEE Trans. Magn.}\ }\textbf {\bibinfo {volume} {40}},\
  \bibinfo {pages} {3443} (\bibinfo {year} {2004})}\BibitemShut {NoStop}%
\bibitem [{\citenamefont {Ellis}\ \emph {et~al.}(2015)\citenamefont {Ellis},
  \citenamefont {Evans}, \citenamefont {Ostler}, \citenamefont {Barker},
  \citenamefont {Atxitia}, \citenamefont {Chubykalo-Fesenko},\ and\
  \citenamefont {Chantrell}}]{Ellis2015_LL}%
  \BibitemOpen
  \bibfield  {author} {\bibinfo {author} {\bibfnamefont {M.~O.~A.}\
  \bibnamefont {Ellis}}, \bibinfo {author} {\bibfnamefont {R.~F.~L.}\
  \bibnamefont {Evans}}, \bibinfo {author} {\bibfnamefont {T.~A.}\ \bibnamefont
  {Ostler}}, \bibinfo {author} {\bibfnamefont {J.}~\bibnamefont {Barker}},
  \bibinfo {author} {\bibfnamefont {U.}~\bibnamefont {Atxitia}}, \bibinfo
  {author} {\bibfnamefont {O.}~\bibnamefont {Chubykalo-Fesenko}}, \ and\
  \bibinfo {author} {\bibfnamefont {R.~W.}\ \bibnamefont {Chantrell}},\ }\href
  {\doibase 10.1063/1.4930971} {\bibfield  {journal} {\bibinfo  {journal} {Low
  Temp. Phys.}\ }\textbf {\bibinfo {volume} {41}},\ \bibinfo {pages} {908}
  (\bibinfo {year} {2015})}\BibitemShut {NoStop}%
\bibitem [{\citenamefont {Evans}\ \emph {et~al.}(2014)\citenamefont {Evans},
  \citenamefont {Fan}, \citenamefont {Chureemart}, \citenamefont {Ostler},
  \citenamefont {Ellis},\ and\ \citenamefont {Chantrell}}]{Evans2014}%
  \BibitemOpen
  \bibfield  {author} {\bibinfo {author} {\bibfnamefont {R.~F.~L.}\
  \bibnamefont {Evans}}, \bibinfo {author} {\bibfnamefont {W.~J.}\ \bibnamefont
  {Fan}}, \bibinfo {author} {\bibfnamefont {P.}~\bibnamefont {Chureemart}},
  \bibinfo {author} {\bibfnamefont {T.~A.}\ \bibnamefont {Ostler}}, \bibinfo
  {author} {\bibfnamefont {M.~O.~A.}\ \bibnamefont {Ellis}}, \ and\ \bibinfo
  {author} {\bibfnamefont {R.~W.}\ \bibnamefont {Chantrell}},\ }\href {\doibase
  10.1088/0953-8984/26/10/103202} {\bibfield  {journal} {\bibinfo  {journal}
  {J. Phys. Condens. Matter}\ }\textbf {\bibinfo {volume} {26}},\ \bibinfo
  {pages} {103202} (\bibinfo {year} {2014})}\BibitemShut {NoStop}%
\bibitem [{\citenamefont {Brown}(1963)}]{Brown1963}%
  \BibitemOpen
  \bibfield  {author} {\bibinfo {author} {\bibfnamefont {W.~F.}\ \bibnamefont
  {Brown}},\ }\href {http://prola.aps.org/abstract/PR/v130/i5/p1677{\_}1}
  {\bibfield  {journal} {\bibinfo  {journal} {Phys. Rev.}\ }\textbf {\bibinfo
  {volume} {130}},\ \bibinfo {pages} {1677} (\bibinfo {year}
  {1963})}\BibitemShut {NoStop}%
\bibitem [{\citenamefont {Pan}\ \emph {et~al.}(2017)\citenamefont {Pan},
  \citenamefont {Chico}, \citenamefont {Delin}, \citenamefont {Bergman},\ and\
  \citenamefont {Bergqvist}}]{Pan2017}%
  \BibitemOpen
  \bibfield  {author} {\bibinfo {author} {\bibfnamefont {F.}~\bibnamefont
  {Pan}}, \bibinfo {author} {\bibfnamefont {J.}~\bibnamefont {Chico}}, \bibinfo
  {author} {\bibfnamefont {A.}~\bibnamefont {Delin}}, \bibinfo {author}
  {\bibfnamefont {A.}~\bibnamefont {Bergman}}, \ and\ \bibinfo {author}
  {\bibfnamefont {L.}~\bibnamefont {Bergqvist}},\ }\href {\doibase
  10.1103/PhysRevB.95.184432} {\bibfield  {journal} {\bibinfo  {journal} {Phys.
  Rev. B}\ }\textbf {\bibinfo {volume} {95}},\ \bibinfo {pages} {184432}
  (\bibinfo {year} {2017})},\ \Eprint {http://arxiv.org/abs/1702.05011}
  {arXiv:1702.05011} \BibitemShut {NoStop}%
\bibitem [{\citenamefont {Mryasov}(2004)}]{Mryasov2004}%
  \BibitemOpen
  \bibfield  {author} {\bibinfo {author} {\bibfnamefont {O.~N.}\ \bibnamefont
  {Mryasov}},\ }\href {\doibase 10.1016/j.jmmm.2003.11.285} {\bibfield
  {journal} {\bibinfo  {journal} {J. Magn. Magn. Mater.}\ }\textbf {\bibinfo
  {volume} {272-276}},\ \bibinfo {pages} {800} (\bibinfo {year}
  {2004})}\BibitemShut {NoStop}%
\bibitem [{\citenamefont {Becker}\ \emph {et~al.}(2014)\citenamefont {Becker},
  \citenamefont {Mosendz}, \citenamefont {Weller}, \citenamefont {Kirilyuk},
  \citenamefont {Maan}, \citenamefont {Christianen}, \citenamefont {Rasing},\
  and\ \citenamefont {Kimel}}]{Becker2014}%
  \BibitemOpen
  \bibfield  {author} {\bibinfo {author} {\bibfnamefont {J.}~\bibnamefont
  {Becker}}, \bibinfo {author} {\bibfnamefont {O.}~\bibnamefont {Mosendz}},
  \bibinfo {author} {\bibfnamefont {D.}~\bibnamefont {Weller}}, \bibinfo
  {author} {\bibfnamefont {A.}~\bibnamefont {Kirilyuk}}, \bibinfo {author}
  {\bibfnamefont {J.~C.}\ \bibnamefont {Maan}}, \bibinfo {author}
  {\bibfnamefont {P.~C.~M.}\ \bibnamefont {Christianen}}, \bibinfo {author}
  {\bibfnamefont {T.}~\bibnamefont {Rasing}}, \ and\ \bibinfo {author}
  {\bibfnamefont {A.~V.}\ \bibnamefont {Kimel}},\ }\href {\doibase
  10.1063/1.4871869} {\bibfield  {journal} {\bibinfo  {journal} {Appl. Phys.
  Lett.}\ }\textbf {\bibinfo {volume} {104}},\ \bibinfo {pages} {152412}
  (\bibinfo {year} {2014})}\BibitemShut {NoStop}%
\bibitem [{\citenamefont {Barker}\ \emph {et~al.}(2010)\citenamefont {Barker},
  \citenamefont {Evans}, \citenamefont {Chantrell}, \citenamefont {Hinzke},\
  and\ \citenamefont {Nowak}}]{Barker2010}%
  \BibitemOpen
  \bibfield  {author} {\bibinfo {author} {\bibfnamefont {J.}~\bibnamefont
  {Barker}}, \bibinfo {author} {\bibfnamefont {R.~F.~L.}\ \bibnamefont
  {Evans}}, \bibinfo {author} {\bibfnamefont {R.~W.}\ \bibnamefont
  {Chantrell}}, \bibinfo {author} {\bibfnamefont {D.}~\bibnamefont {Hinzke}}, \
  and\ \bibinfo {author} {\bibfnamefont {U.}~\bibnamefont {Nowak}},\ }\href
  {\doibase 10.1063/1.3515928} {\bibfield  {journal} {\bibinfo  {journal}
  {Appl. Phys. Lett.}\ }\textbf {\bibinfo {volume} {97}},\ \bibinfo {pages}
  {192504} (\bibinfo {year} {2010})}\BibitemShut {NoStop}%
\bibitem [{\citenamefont {Antoniak}\ \emph {et~al.}(2010)\citenamefont
  {Antoniak}, \citenamefont {Lindner}, \citenamefont {Fauth}, \citenamefont
  {Thiele}, \citenamefont {Min{\'{a}}r}, \citenamefont {Mankovsky},
  \citenamefont {Ebert}, \citenamefont {Wende},\ and\ \citenamefont
  {Farle}}]{Antoniak2010}%
  \BibitemOpen
  \bibfield  {author} {\bibinfo {author} {\bibfnamefont {C.}~\bibnamefont
  {Antoniak}}, \bibinfo {author} {\bibfnamefont {J.}~\bibnamefont {Lindner}},
  \bibinfo {author} {\bibfnamefont {K.}~\bibnamefont {Fauth}}, \bibinfo
  {author} {\bibfnamefont {J.-U.}\ \bibnamefont {Thiele}}, \bibinfo {author}
  {\bibfnamefont {J.}~\bibnamefont {Min{\'{a}}r}}, \bibinfo {author}
  {\bibfnamefont {S.}~\bibnamefont {Mankovsky}}, \bibinfo {author}
  {\bibfnamefont {H.}~\bibnamefont {Ebert}}, \bibinfo {author} {\bibfnamefont
  {H.}~\bibnamefont {Wende}}, \ and\ \bibinfo {author} {\bibfnamefont
  {M.}~\bibnamefont {Farle}},\ }\href {\doibase 10.1103/PhysRevB.82.064403}
  {\bibfield  {journal} {\bibinfo  {journal} {Phys. Rev. B}\ }\textbf {\bibinfo
  {volume} {82}},\ \bibinfo {pages} {064403} (\bibinfo {year}
  {2010})}\BibitemShut {NoStop}%
\bibitem [{\citenamefont {Kubo}(1966)}]{Kubo1966}%
  \BibitemOpen
  \bibfield  {author} {\bibinfo {author} {\bibfnamefont {R.}~\bibnamefont
  {Kubo}},\ }\href {\doibase 10.1088/0034-4885/29/1/306} {\bibfield  {journal}
  {\bibinfo  {journal} {Reports Prog. Phys.}\ }\textbf {\bibinfo {volume}
  {29}},\ \bibinfo {pages} {306} (\bibinfo {year} {1966})}\BibitemShut
  {NoStop}%
\bibitem [{\citenamefont {Butera}(2006)}]{Butera2006}%
  \BibitemOpen
  \bibfield  {author} {\bibinfo {author} {\bibfnamefont {A.}~\bibnamefont
  {Butera}},\ }\href {\doibase 10.1140/epjb/e2006-00296-4} {\bibfield
  {journal} {\bibinfo  {journal} {Eur. Phys. J. B}\ }\textbf {\bibinfo {volume}
  {52}},\ \bibinfo {pages} {297} (\bibinfo {year} {2006})}\BibitemShut
  {NoStop}%
\bibitem [{\citenamefont {Evans}\ \emph {et~al.}(2012)\citenamefont {Evans},
  \citenamefont {Hinzke}, \citenamefont {Atxitia}, \citenamefont {Nowak},
  \citenamefont {Chantrell},\ and\ \citenamefont
  {Chubykalo-Fesenko}}]{Evans2012}%
  \BibitemOpen
  \bibfield  {author} {\bibinfo {author} {\bibfnamefont {R.~F.~L.}\
  \bibnamefont {Evans}}, \bibinfo {author} {\bibfnamefont {D.}~\bibnamefont
  {Hinzke}}, \bibinfo {author} {\bibfnamefont {U.}~\bibnamefont {Atxitia}},
  \bibinfo {author} {\bibfnamefont {U.}~\bibnamefont {Nowak}}, \bibinfo
  {author} {\bibfnamefont {R.~W.}\ \bibnamefont {Chantrell}}, \ and\ \bibinfo
  {author} {\bibfnamefont {O.}~\bibnamefont {Chubykalo-Fesenko}},\ }\href
  {\doibase 10.1103/PhysRevB.85.014433} {\bibfield  {journal} {\bibinfo
  {journal} {Phys. Rev. B}\ }\textbf {\bibinfo {volume} {85}},\ \bibinfo
  {pages} {014433} (\bibinfo {year} {2012})}\BibitemShut {NoStop}%
\end{thebibliography}%

\end{document}